\documentclass{aa}
\usepackage[switch]{lineno}
\usepackage{natbib}
\usepackage{graphicx}
\usepackage{subfigure}
\usepackage{txfonts}
\usepackage{hyperref}
\usepackage{color}

\begin{document}
\title{The evolutionary pathways of disk galaxies with different sizes}
\titlerunning{The evolutionary pathways of disk galaxies with different sizes}
\authorrunning{Ma, Du, et al.}
\author{Hong-Chuan Ma\inst{1}
\and 
Min Du\inst{1}\fnmsep\thanks{Min Du and Hong-Chuan Ma have equal contributions to this work. Any response should be directed to Min Du (e-mail:dumin@xmu.edu.cn)}
\and
Luis C. Ho\inst{2, 3}
\and 
Ming-Jie Sheng\inst{1}
\and
Shihong Liao\inst{4}}
\institute{
            $^1$ Department of Astronomy, Xiamen University, Xiamen, Fujian 361005, China \\
            $^2$ Kavli Institute for Astronomy and Astrophysics, Peking University, Beijing 100871, China \\
            $^3$ Department of Astronomy, School of Physics, Peking University, Beijing 100871, China\\
            $^4$ Key Laboratory for Computational Astrophysics, National Astronomical Observatories, Chinese Academy of Sciences, Beĳing 100101, China
             }
\date{Received ???; accepted ???}

\abstract 
{}
{This research delves into the complex interaction between disk galaxies and their host dark matter halos. It specifically focuses on scenarios with minimal external influences or ``nurture'' such as mergers and substantial tidal interactions. The study uncovers the varied evolutionary paths of disk galaxies with different sizes, shaped by the initial conditions of their parent dark matter halos and subsequent internal processes, namely the ``nature'' of the galaxies.}
{From the TNG50 simulation, a sample of 836 central disk galaxies with tiny stellar halos is chosen to study the inherent evolution of galaxies driven by nature. These galaxies are classified as compact, normal, or extended by referencing their locations on the mass-size ($M_\star-R_{\rm 1/2}$) diagram. Scaling relations are then established to measure the correlations driven by internal mechanisms.}
{This research demonstrates the distinctive evolutionary pathways of galaxies with different sizes in IllustrisTNG simulations, primarily driven by nature. It is confirmed that disk galaxies inherit the angular momentum of their parent dark matter halos. More compact galaxies form earlier within halos possessing lower specific angular momentum through heightened star formation during the early phase at redshifts above 2. During the later phase, the size of extended galaxies experiences more pronounced growth by accreting gas with high angular momentum. Additionally, we reveal that many key characteristics of galaxies are linked to their mass and size: (1) compact galaxies tend to exhibit higher metal content, proportional to the potential well $\frac{M_\star}{R_{\rm 1/2}}$, (2) compact galaxies host more massive bulges and black holes, and higher central concentration. Furthermore, our analysis indicates that galaxies of all types continue to actively engage in star formation, with no evident signs of quenching attributed to their varying sizes and angular momenta.}
{}

\maketitle
%
\section{Introduction}
\label{sec:intro}

    It is essential to comprehend the physical reasons behind the size of galaxies to grasp the process of galaxy formation and evolution. Galaxies span a large range of sizes in both observational and simulation results, with effective radii or half-mass radii ranging from 0.1 kpc to 10 kpc \citep[e.g.][]{Shen2003,Bernardi2010,Lange2015, Pillepich2019}. In the standard picture of galaxy formation, galaxies acquire their angular momenta from the dark matter halos they reside in, which originate from the cosmological tidal torque field \citep{Hoyle1951,Peebles1969,Doroshkevich1970,White1984}. This analytical model has been extensively applied to forecast the angular momentum and size of galaxies assuming the angular momentum remains conserved during the formation of galaxies that exhibit a roughly exponential surface density distribution along their radius, maintaining centrifugal equilibrium \citep{Fall1980, Dalcanton1997, Mo1998, Dutton2007, Somerville2008, Guo2011, Benson2012}. However, this picture is still under active debate. Various factors could potentially influence the size of galaxies, such as the conservation efficiency of angular momentum \citep{Danovich2015}, the mass fraction of spheroidal structures, the alteration of the exponential surface density profile as a result of stellar migration in disks \citep[e.g.][]{Debattista2006, Roskar2012, Berrier&Sellwood2015}, the faster collapse of material with low angular momentum, the so-called biased collapse \citep{Shi2017, Posti2018b}, the stellar and active galactic nucleus (AGN) feedback \citep{Somerville2015, Naab2017}, and even the different physical origins of the exponential surface density \citep{Yoshii1989,Ferguson2001,Elmegreen2013,Wang2022}. 

    There is evidence suggesting that the size of galaxies is largely determined by their parent dark matter halos, although some studies found weak/no correlation. \cite{Zavala2016} and \cite{Lagos2017} found a notable connection between the specific angular momentum $j\equiv J/M$ evolution of dark matter and the baryonic components of galaxies in EAGLE simulations. \citet{Grand2017} showed that galaxy sizes correlate with the spin parameter of dark matter halos using 30 galaxies from the Auriga zoom-in simulations. A consistent result is obtained in IllustrisTNG \citep[e.g.][]{Rodriguez-Gomez2022, Yang2023}. \citet{Cadiou2022} suggested that the stellar angular momentum and size of galaxies can be affected by their cosmological initial conditions via examining three zoom-in simulations. \cite{Desmond2017} found a weak correlation between galaxy size and the host halo spin parameter in the EAGLE simulation. \cite{Zanisi2020} argued that the scatter in the galaxy-halo size relation for late-type galaxies could be explained by the scatter in stellar angular momentum rather than the halo spin parameter. \cite{Varela-Lavin2023} reported a correlation between asymmetry and spin using the IllutrisTNG(TNG)-50 simulations. \citet{Jiang2019} and \citet{Liang2024} suggested that the central concentration of dark matter halos may also affect the size of galaxies. 

    Recently, \citet{Du2022} and \citet{Du2024} presented a clear picture about the correlations of stellar mass $M_\star$, size, and specific angular momentum of stars $j_\star$ using disk galaxies in the TNG simulations \citep{Marinacci2018, Nelson2018a, Nelson2019a, Naiman2018, Pillepich2018a, Pillepich2019, Springel2018}. \cite{Du2022} showed that the $j_\star \propto M_\star ^{0.55}$ relation in TNG is consistent with observations \citep{Fall1980, Romanowsky&Fall2012,Posti2018a, DiTeodoro2021, Mancera-Pina2021, Hardwick2022a, Rodriguez-Gomez2022}. \citet{Du2022} further showed that the formation of disk galaxies in the TNG simulations aligns with the framework proposed by \cite{Mo1998}. It suggested that galaxies indeed inherit the angular momentum of their parent dark matter halos, but adjustments possibly due to baryonic processes are non-negligible. The biased collapse does not have a distinct impact in the local Universe. \citet{Du2024} further confirmed the existence of a strong relationship between $j_\star$--$M_\star$ and galaxy size in TNG galaxies that are shaped by both initial conditions and internal processes, i.e., the ``nature''. The intrinsic variation in $j_\star$ accounts for the mass-size relationship and its significant variability. This finding emphasizes the primary influence of nature on shaping the overall $j_\star$--$M_\star$ and mass-size relations in disk galaxies, while other effects play a relatively minor role. See more discussion about the ``nature-nurture'' dichotomy in \citet[][see cartoon Figure 26]{Du2021}.

    The different sizes of galaxies possibly lead to distinct evolution processes. In a study by \citet[][Figure 16 there]{Du2021}, it was demonstrated that galaxies dominated by bulges and disks, with minimal external influences, evolve along separate paths. Bulge-dominated galaxies follow a compact pathway that consists of two phases: first, a compact phase where mass increases significantly while size remains relatively constant, leading to the formation of bulges; and second, a phase where size increases considerably while mass growth is relatively limited, resulting in the development of disk-like structures. On the other hand, disk-dominated galaxies gradually form disky structures in a more prolonged manner. Therefore, there is a clear connection between the size of galaxies and the recycling of gas in the circumgalactic region, as well as their star formation history. This relationship may further influence the evolution of galaxies in many aspects, e.g., the metal enrichment and the growth of central supermassive black holes \citep{Sanchez-Menguiano2024,Vaughan2022, Barone2020, Barone2018, D'Eugenio2018, Krajnovi'c2018, Grand2017, Chang2017}. Therefore, it is crucial to comprehend the underlying physical reasons behind the mass-size relationship and what role it will play in the formation and evolution of galaxies.

    \begin{figure}[htb]
            \centering
            \includegraphics[width=0.45\textwidth]{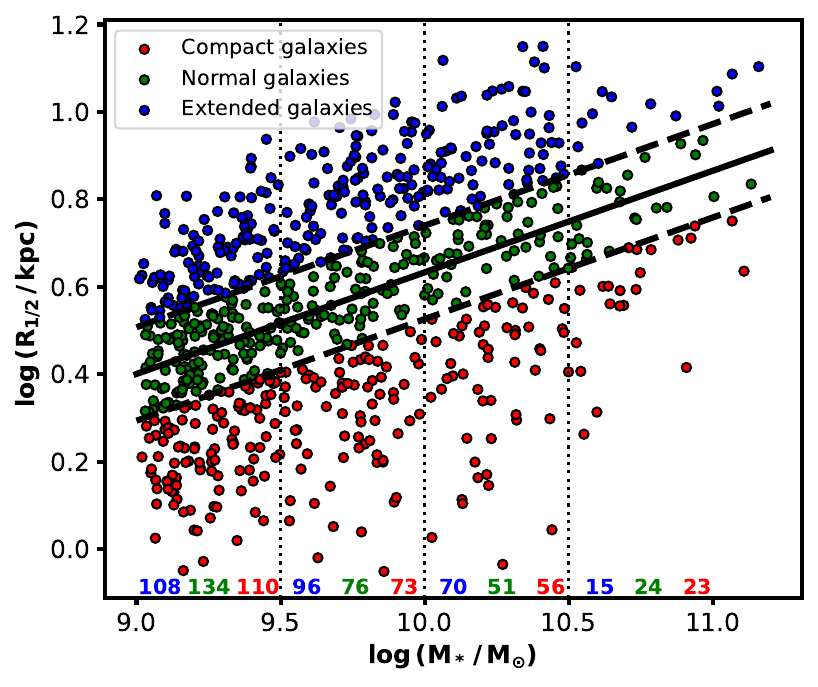}
            \caption{The stellar mass-size relation for the central disk galaxies with tiny stellar halos we selected in the TNG50 simulation at $z = 0$. A linear fit of the mass-size correlation is represented by the solid black line. We categorize galaxies based on their deviation within $\pm 0.5 \sigma$ from the linear fit, as the dashed lines indicate. Galaxies falling within the range of $-0.5\sigma$ to $0.5\sigma$ are identified as 'normal galaxies' (green dots), those exceeding $0.5\sigma$ are labeled as 'extended galaxies' (blue dots), and those falling below $-0.5\sigma$ are designated as 'compact galaxies' (red dots). The dotted lines partition the galaxies into four mass bins (M1-M4) encompassing $10^{9-9.5} M_{\odot}$, $10^{9.5-10} M_{\odot}$, $10^{10-10.5} M_{\odot}$, and $10^{10.5-11.5} M_{\odot}$. The bottom of this figure provides the count of galaxies of each type within M1-M4.}
            \label{Fig 1}
    \end{figure}
        \begin{figure*}[htb]
            \centering
            \includegraphics[width=1\textwidth]{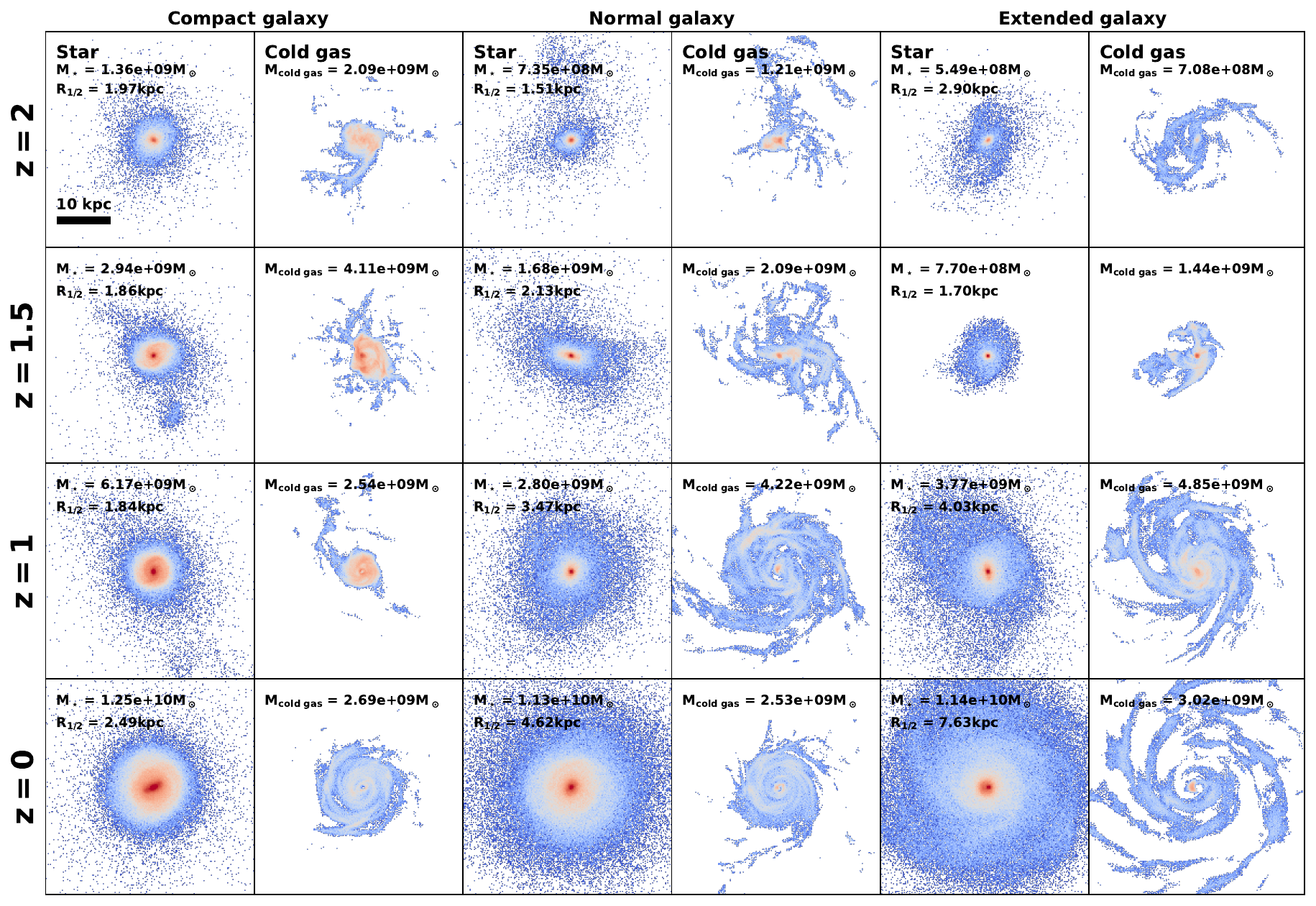}
            \caption{The surface density distributions of stars and cold gas in the face-on view of compact (left), normal (middle), and extended (right) galaxies at $z = 0, 1, 1.5, 2$. From top to bottom, we show the evolution of stellar and cold gas. Their stellar mass $M_\star$ and 2D half-mass radius $R_{1/2}$ are given on the left-hand corner of each panel.}
            \label{Fig 2}
        \end{figure*}

        \begin{figure*}
            \centering
            \includegraphics[scale=0.5]{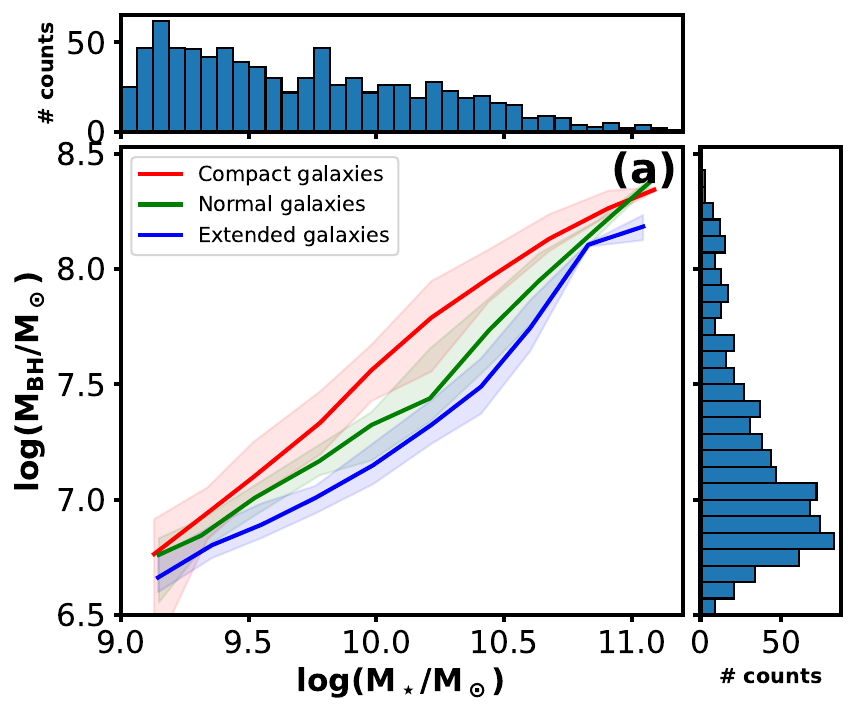}
            \hspace{1in}
            \includegraphics[scale=0.5]{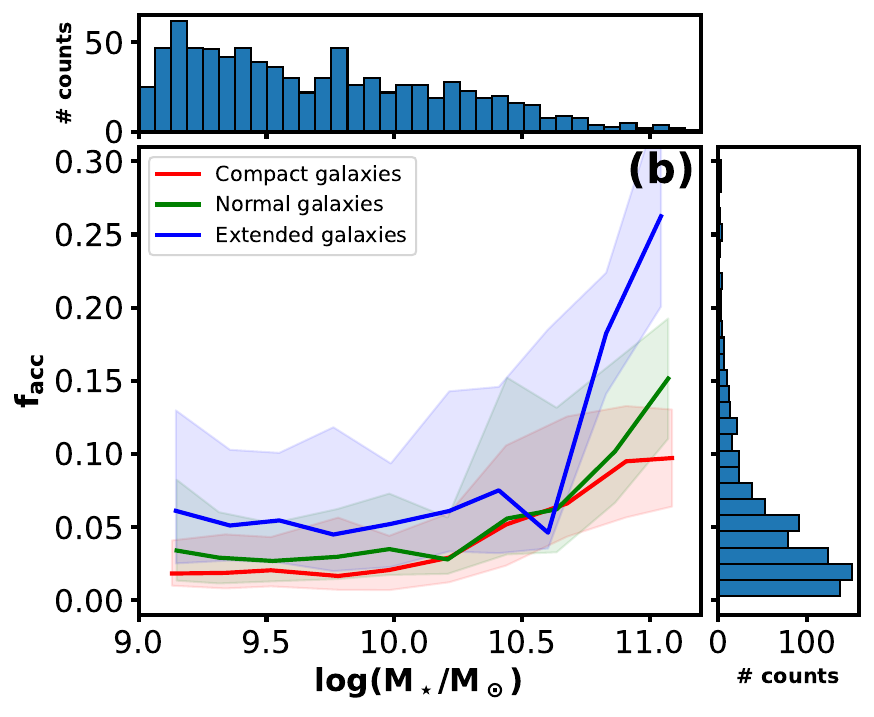}
            \caption{Properties of galaxies in our sample at $z = 0$. \textit{Panel \rm (a)}: Black hole mass as a function of stellar mass for the compact, normal, and extended galaxies. The top and right sub-panels show the histograms computed from all galaxies, respectively. \textit{Panel \rm (b)}: Ex situ mass fraction $f_{\rm acc}$ as a function of the stellar mass, and the histogram distributions.}
            \label{Fig 3}
        \end{figure*}
    
    In this study, we investigate the evolutionary paths of galaxies with different sizes in the TNG simulations. We select galaxies that have been minimally impacted by external processes (e.g., mergers). In practice, we define these as galaxies that have a small fraction of their stellar mass residing in their halo component, which \citet{Du2021} have shown to be a robust tracer of the merger history of the galaxy. We analyze how a galaxy's mass and size influence its other characteristics through internal processes, which represent the natural progression of evolution in \citet{Du2021}. We provide a brief introduction of the simulation and the selection criteria of galaxy samples in Section \ref{sec:2}. The two-phase evolution history of galaxies is illustrated in Section \ref{sec:3}. Section \ref{sec:4} provides a detailed analysis of the relationship between scatters in the mass-size relation and other galaxy properties. In Section \ref{sec:5}, we summarize the results.

\section{Simulation and sample selection} \label{sec:2}

    \subsection{The IllustrisTNG project}

        TNG is a suite of gravo-magnetohydrodynamical simulations running with the moving-mesh code \small{AREPO} \citep{Springel2010}, and followed the standard $\Lambda$CDM model with parameters based on the results of \cite{Planck2016}: $\Omega_m =0.3089$, $\Omega_{\Lambda}=0.6911$, $\Omega_b=0.0486$, $h=0.6774$, $\sigma_8=0.8159$, $n_s=0.9667$. The TNG project is the successor of the original Illustris simulation \citep{Vogelsberger2014,Genel2014,Nelson2015, Sijacki2015} via including an updated galaxy formation model, such as physical model improvement (e.g., black hole low-state feedback changes from `radio' bubbles to black hole-driven wind), numerical optimization, and treatment of cosmic magnetism \citep{Weinberger2017,Pillepich2018a,Nelson2019a}. It is composed of three fiducial runs, namely TNG50-1, TNG100-1, and TNG300-1, that use different simulation boxes and resolutions.
        
        In this work, we use the TNG50-1 (hereafter TNG50) simulation \citep{Nelson2019,Nelson2019a,Pillepich2019}. It includes $2160^3$ baryonic elements and $2160^3$ dark matter particles in a periodic box with side length $\sim 50 \, \text{comoving Mpc}$, whose mass resolutions reach $8.41 \times 10^4 M_\odot$ and $4.57 \times 10^5 M_\odot$, respectively. The gravitational softening length of both dark matter and stellar particles is $0.29 \, \text{kpc}$ at $z = 0$, and the minimum gas softening length reaches $0.074 \, \text{comoving kpc}$. Halos and subhalos are identified and characterized using the friends-of-friends \citep{Davis1985} and \small{SUBFIND} algorithm \citep{Springel2001}, respectively. We define that galaxies are composed of all gas and stars in their host subhalo. Unless otherwise noted, we measure all quantities by using all particles of galaxies. The progenitors of galaxies are traced back along their main progenitor branches in the merger trees.

    \subsection{Sample selection and classification of disk galaxies: compact, normal, and extended}\label{sec:2.2}

        \begin{figure*}[htb]
            \centering
            \includegraphics[width=1\textwidth]{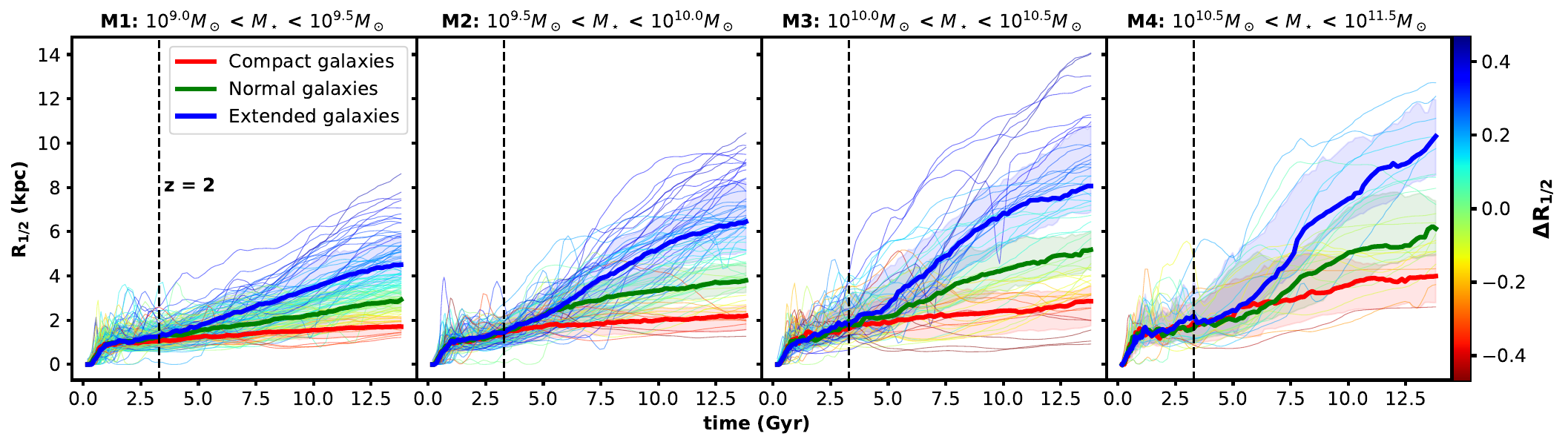}
            \caption{The evolutionary trajectory of the galaxy we have chosen in terms of its half-mass radius $R_{1/2}$ growth. Each solid line corresponds to a specific galaxy, color-coded according to $\Delta R_{1/2}$. The bold solid lines and shaded areas indicate the median and the $1\sigma$ variation for compact (red), normal (green), and extended (blue) galaxies. The galaxies are categorized into M1-M4 from left to right. Vertical dashed lines are the time at redshift $z = 2$.}
            \label{Fig 4}
        \end{figure*}

        \begin{figure*}[htb]
            \centering
            \includegraphics[width=0.9\textwidth]{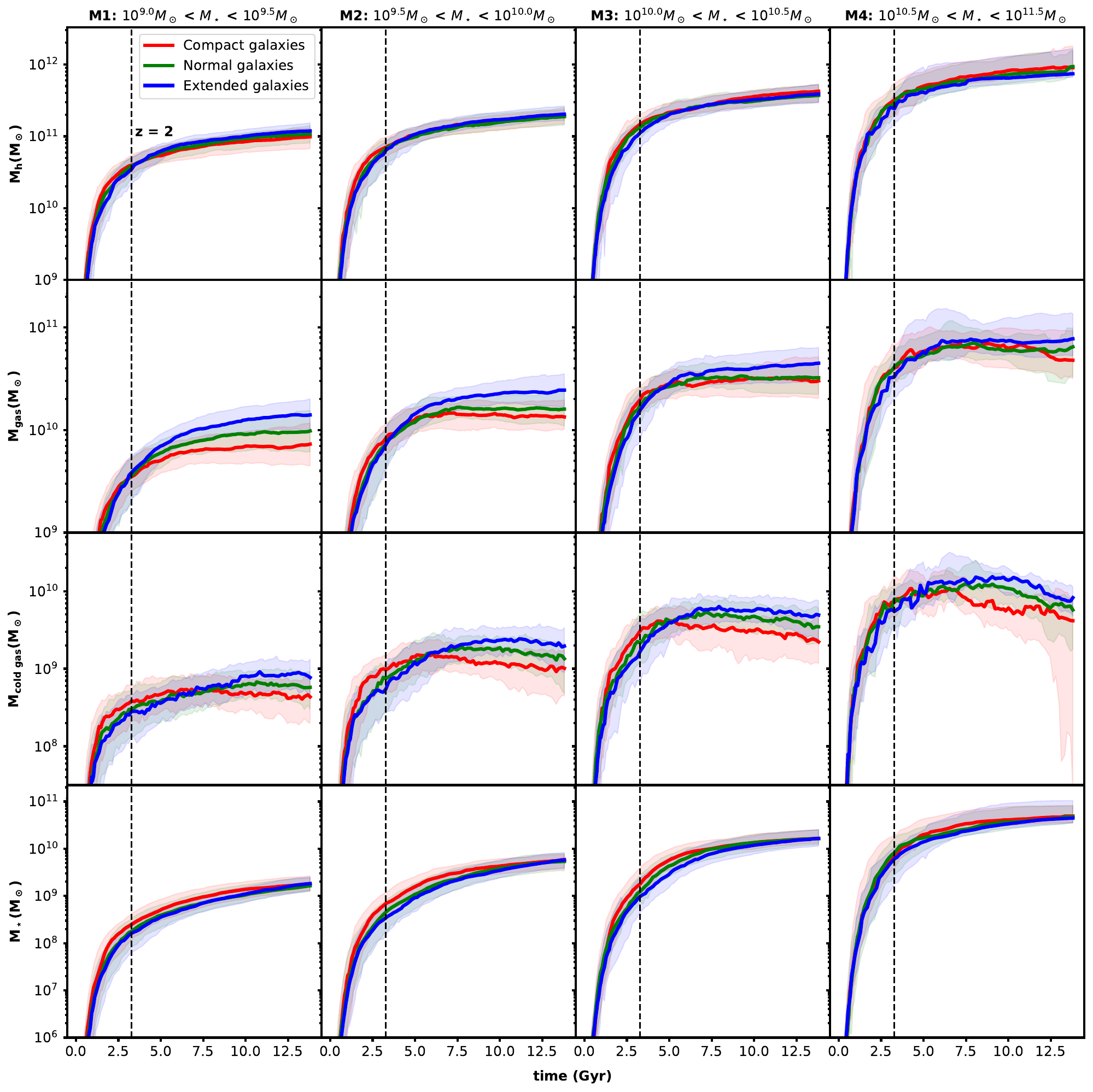}
            \caption{The mass growth history of dark matter halo, gas, cold gas, and stars (from top to bottom) in compact (red), normal (green), and extended (blue) galaxies we selected. The solid profiles and shaded regions represent the median and $1\sigma$ envelope. From left to right galaxies are divided into M1-M4.}
            \label{Fig 5}
        \end{figure*}

        \begin{figure*}[htb]
            \centering
            \includegraphics[width=0.9\textwidth]{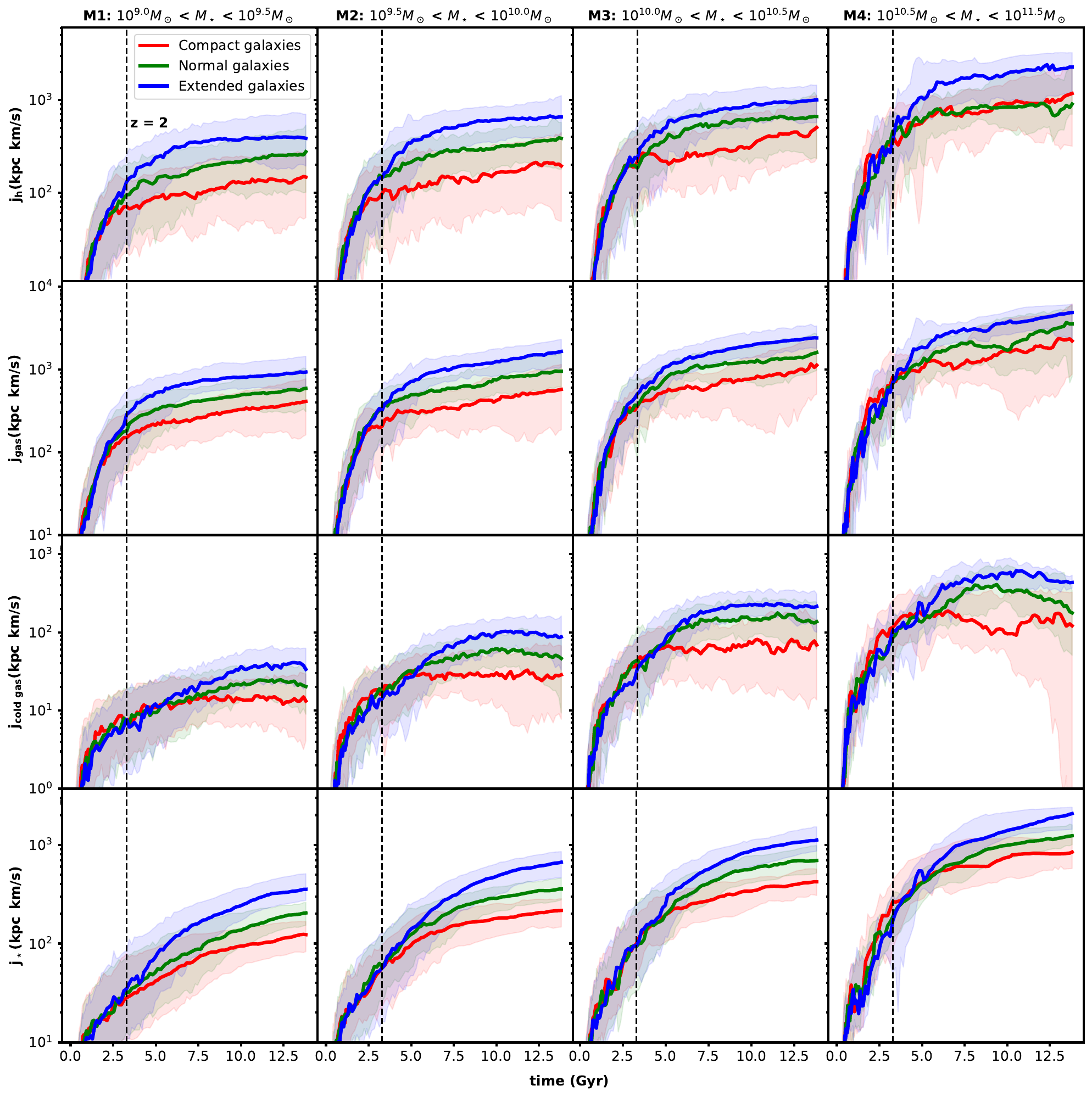}
            \caption{The angular momentum growth history of dark matter halos, gas, cold gas, and stars (from top to bottom) in compact (red), normal (green), and extended (blue) galaxies we select. The solid profiles and shaded regions represent the median and $1\sigma$ envelope. From left to right, galaxies are divided into M1-M4.}
            \label{Fig 6}
        \end{figure*}

        In this study, we aim to examine how galaxies of varying sizes evolve in the case that internal processes dominate. We thus select central galaxies with tiny kinematically defined stellar halos to minimize the influence of the external processes and environment. The specific selection criteria are outlined below. We use galaxies with stellar masses larger than $10^9 M_{\odot}$ that are able to reproduce well the overall kinematic properties of galaxies \citep{Pillepich2019}. Next, We adopt the dynamical decomposition results from \cite{Du2020,Du2021}\footnote{The data is publicly accessible at https://www.tng-project.org/data/docs/specifications/\#sec5m}, which were obtained by employing automatic Gaussian Mixture Models (auto-GMM) to cluster circularity, binding energy, and non-azimuthal angular momentum in an accurate, efficient, and unsupervised way \citep{Du2019}. This method separates galaxies into distinct structures: cold disks, warm disks, disky bulges, bulges, and stellar halos. Notably, the stellar particles of stellar halos are primarily characterized by large random motions and weak gravitational binding, primarily induced by external physical processes. Therefore, we select galaxies with tiny stellar halos whose mass fraction is $f_{\rm halo}<0.2$, aiming to minimize the influence of mergers and strong tidal interactions. We have confirmed that nearly all galaxies we selected are disk galaxies characterized by $\kappa_{\rm rot}>0.5$. $\kappa_{\rm rot}$ is defined as
        \begin{equation}
        \kappa_{\rm rot} = \frac{K_{\rm rot}}{K} = \frac{1}{K}\sum\frac{1}{2}m(\frac{j_z}{R})^2, 
        \end{equation}
        where $j_{\rm z}$ denotes the specific angular momentum along the minor-axis $z$-axis of each galaxy, $R$ is the projected radius and $K$ is the total kinetic energy \citep{Sales2012}. A few galaxies with $\kappa_{\rm rot}<0.5$ in the sample selected by $f_{\rm halo} < 0.2$ are excluded. In the end, 836 galaxies are left.

        Figure \ref{Fig 1} shows the stellar mass-size relation of the TNG50 galaxy sample chosen at $z=0$. Here the galaxy size corresponds to the half-mass radius measured in the face-on view $R_{1/2}$. In general, disk galaxies have a wide range of sizes. We find $R_{1/2}$ can vary by up to a factor of approximately $10$ for galaxies with the same mass in Figure \ref{Fig 1}. \citet{Du2024} verified that the size of galaxies largely depends on the angular momentum under the exponential hypothesis using the same sample we selected here. We thus classify galaxies into compact, normal, and extended galaxies in this work to investigate the different evolutionary pathways in galaxies with different sizes. It's worth noting that the galaxies we have selected are distributed continuously. The best-fit mass-size relation is obtained from linear regression 
        \begin{equation}
            \rm log(R_{1/2, fit}/kpc) = 0.23\,log(M_\star/M_\odot) - 1.69.
        \end{equation}
        It gives a large scatter $\sigma=0.21$ of log ($R_{\rm 1/2}$ / kpc). We then quantify the deviation from the predicted size log$R_{\rm 1/2,fit}$ of such a mass-size relation using
        \begin{equation}\label{equ:3}
            \Delta R_{1/2} = \rm log(R_{1/2}) - log(R_{\rm 1/2,fit}).
        \end{equation}
        Galaxies with $\Delta R_{1/2} > 0.5 \sigma$ are classified as extended galaxies, while those with $\Delta R_{1/2} < -0.5 \sigma$ are classified as compact galaxies. Galaxies with $-0.5 \sigma < \Delta R_{1/2} < 0.5 \sigma$ fall into the category of normal galaxies. Additionally, we segmented the mass range into four bins, \textbf{M1}: $10^9 M_{\odot} < M_{\star} < 10^{9.5} M_{\odot}$, \textbf{M2}: $10^{9.5} M_{\odot} < M_{\star} < 10^{10} M_{\odot}$, \textbf{M3}: $10^{10} M_{\odot} < M_{\star} < 10^{10.5} M_{\odot}$ and \textbf{M4}: $M_{\star} > 10^{10.5} M_{\odot}$, Due to the rarity of galaxies with stellar mass exceeding $10^{11} M_{\odot}$ within our sample, we have included them in the category of \textbf{M4}. The number of each type of galaxies is provided at the bottom of Figure \ref{Fig 1}.

        Figure \ref{Fig 2} gives examples of a compact (ID 631677), a normal (ID 621707), and an extended (ID 619081) galaxy from our selected sample. The panels show the surface density distributions of face-on projected cold gas and stars at $z =$ 0, 1, 1.5, and 2 from bottom to top. Here, we take the star-forming gas cells as cold, and the rest as hot. The stellar mass $M_{\rm \star}$, cold gas mass $M_{\rm cold\,gas}$, and size $R_{1/2}$ of each galaxy are displayed at the top left-hand corner. It is clear that the evolution of the galaxies we selected here is dominated by internal processes. Compact galaxies form from a more compact cold gas disk.

        \subsection{Selected galaxies minimally influenced by external processes and AGN feedback quenching}

            It is well known that the AGN feedback can quench galaxies. Hence, we investigate the black hole masses across galaxies of varying sizes to ascertain the presence of AGN feedback. Figure \ref{Fig 3} shows the properties of the sample and variations in properties among different galaxy types at $z=0$. Panel (a) of Figure \ref{Fig 3} shows the $M_\star$--$M_{\rm BH}$ relation for compact, normal, and extended galaxies. The right sub-panel shows the histogram of black hole masses, with the majority of galaxies having black hole masses below $10^8 M_{\odot}$. It is lower than the threshold for turning on the radio (kinetic) mode AGN feedback that starts to play an important role in quenching star formation in the TNG simulation \citep{Weinberger2017, Weinberger2018}. This suggests that our selected galaxies have not been strongly quenched by AGN feedback.

            We further examine to what extent the sample selected based on $f_{\rm halo} < 0.2$ avoided the influence of strong interactions and mergers. We adopt the stellar assembly history supplementary catalog provided on the TNG website \citep{Rodriguez-Gomez2015, Rodriguez-Gomez2016, Rodriguez-Gomez2017}\footnote{The data are publicly accessible at https://www.tng-project.org/data/docs/specifications/\#sec5g}. Panel (b) of Figure \ref{Fig 3} shows the ex-situ stellar mass fraction $f_{\rm acc}$ as a function of the stellar mass for compact, normal, and extended galaxies. More extended galaxies tend to have a slightly larger $f_{\rm acc}$ \citep{Zhu2023}.  It is reasonable that ex-situ stars are likely to move on orbits with a large radius and ellipticity \citep[e.g.][]{Naab2009,Dubois2016}. It could result in the enlargement of galaxies.  The right histogram of the panel (b) shows that $f_{\rm acc}$ is generally less than 0.1. The effect of external processes thus is small. $f_{\rm halo}$ is convenient to eliminate the need to account for variations in the details of external processes. And it is measurable in edge-on galaxies using either kinematic \citep{Yang2022} or morphological \citep[e.g.,][]{Gadotti2012} methods. The sample we select is able to unveil the innate evolutionary pathway of galaxies primarily depending on their size and mass.

\section{The evolutionary pathways of galaxies with different sizes} \label{sec:3}

    \cite{Du2022} and \cite{Du2024} have verified that the size of galaxies is largely determined by the inherent scatter of specific angular momentum of their host dark matter halos $j_{\rm h}$. They used an identical galaxy sample characterized by small kinematic stellar halos with a mass ratio $f_{\rm halo} < 0.2$. Such galaxies explain well the observed $j_\star$--$M_\star$ relation and the mass-size relation in the local Universe. Their results suggested that the scatter of the mass-size relations at $z=0$ originates from $j_\star$. However, the $j_\star$, $M_\star$, and $R_{1/2}$ of galaxies are the outcomes of a series of highly complex and nonlinear baryonic physical processes. Therefore, understanding the evolution of the mass-size relation and its scatter requires tracing the progression from dark matter halos to gas, then to cold gas, and finally to stars. This section aims to uncover the growth history of the size, mass, and specific angular momentum of dark matter halos, gas (including cold gas), and stars, thus elucidating their correlations.
    
    \subsection{Early Phase Formation of Galaxies: All Galaxies with Similar Specific Angular Momentum and Size but Compact Galaxies with Slightly Higher Mass}\label{sec 3.1}

    \begin{figure}[htb]
            \centering
            \includegraphics[width=0.45\textwidth]{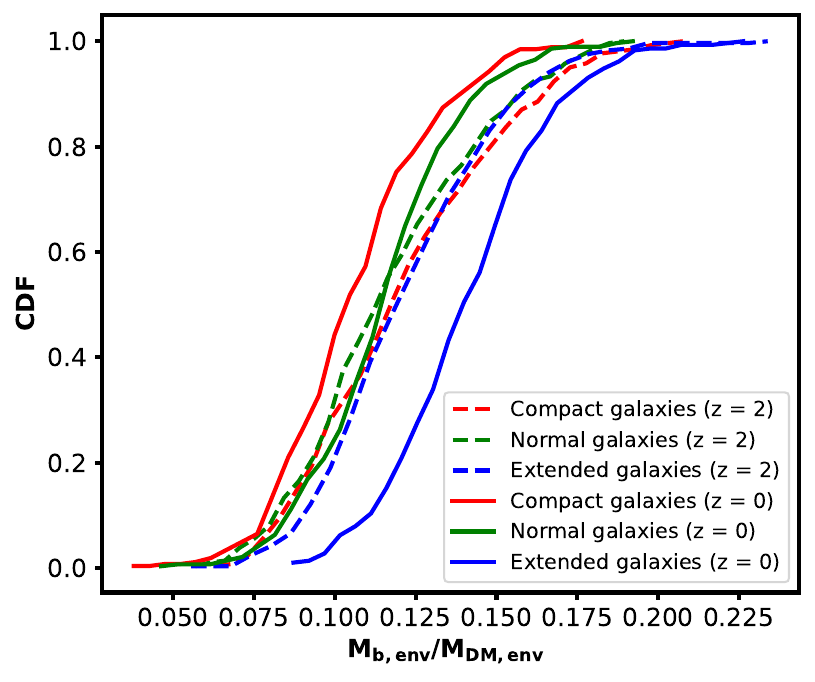}
            \caption{The cumulative distribution of baryonic to dark matter mass ratio of the group or cluster environment that compact (red), normal (green), and extend (blue) galaxies belong to at $z = 2$ (dash) and $z = 0$ (solid).}
            \label{Fig 7}
    \end{figure}

        In this work, we regard all processes at $z > 2$ as the early-phase evolution of galaxies and $z < 2$ as the late-phase evolution. Every profile color-coded by $\Delta R_{1/2}$ shows the evolution of one galaxy in Figure \ref{Fig 4}. The shaded areas represent the 1$\sigma$ boundaries of compact (red), normal (green), and extended (blue) galaxies classified in Section \ref{sec:2.2}. Subsequent figures will not display individual galaxy profiles. Instead, the 1$\sigma$ envelopes will be used to emphasize distinctions among galaxy types. In the early phase, there's no significant difference in the size $R_{1/2}$ among the three types of galaxies. Furthermore, all galaxies exhibit comparable $R_{1/2}$ varying roughly from 0.5 kpc to 2 kpc, with a median size of about 1 kpc, as shown in Figure \ref{Fig 4}. Recently, \cite{Ormerod2024} reported result that galaxies with stellar mass greater than $10^{9.5} M_\odot$ observed at $z > 3$ by JWST have a similar size $\sim$ 1 kpc. This result is consistent with the TNG galaxies we selected here, assuming the mass-to-light ratios of galaxies do not vary significantly along the radius.
    
        We further show the evolution of mass and angular momentum in Figures \ref{Fig 5} and \ref{Fig 6}. From top to bottom panels, we trace the growth history of dark matter, total gas, cold gas, and stars in selected galaxies. In the early phase, compact galaxies have a negligible difference in dark matter and total gas mass compared to normal and extended counterparts, shown in the first and second rows of Figure \ref{Fig 5}. However, compact galaxies exhibit slightly more massive stellar components, possibly due to their faster growth via the accretion of more cold gas, as shown in the third row of Figure \ref{Fig 5}. Notably, dark matter, gas (cold gas), and stars have similar and low specific angular momentum, as shown in Figure \ref{Fig 6}. It is clear that the formation of more compact galaxies in the early phase is attributed to the presence of a greater amount of low angular momentum gas, as proposed by \cite{Du2024}. Therefore, this finding aligns with the concept of biased collapse \citep[e.g.][]{van-den-Bosch1998, Shi2017}, suggesting that gas with low specific angular momentum cools down more rapidly. The similar sizes of compact, normal, and extended galaxies can be attributed to the maintenance of similar specific angular momentum throughout the process from dark matter halos to galaxies. Consequently, compact galaxies have slightly more cold gas of low $j_{\rm cold\,gas}$, leading to the formation of somewhat more massive stellar components compared to their more extended counterparts, as shown in the bottom panels of Figure \ref{Fig 5}.
    
        Overall, all galaxies have similar specific angular momentum (e.g., $j_{\rm h}$, $j_{\rm gas}$, $j_{\rm cold\,gas}$, and $j_\star$) and size, but compact galaxies have slightly higher stellar mass. Consequently, the slope of the mass-size relation remains flat at high redshifts. This finding aligns with the mass-size relation of star-forming galaxies derived by \cite{Genel2018} using TNG100 at $z = 2$. It is important to highlight that the early-phase evolution of galaxies likely determines their bulge growth. By redshift $z\sim 2$, galaxies have accumulated approximately 10 percent of their total stellar masses at $z = 0$. It is likely to contribute to the early formation of bulges due to the accretion of low $j_{\rm cold\,gas}$. Hence, it can be concluded that the formation of galaxies with different sizes follows a similar pattern in the early phase and may have a limited impact on determining the overall properties of galaxies.
        
    \subsection{Late Phase Secular Evolution of Galaxies: Gradual Assembly of Extended Disks in Cases of High Angular Momentum}\label{sec 3.2}

        In the late phase $z<2$, the discrepancy of size between compact and extended galaxies increases significantly (Figure \ref{Fig 4}), while in the meanwhile the difference in their stellar masses decreases gradually shown in the bottom panels of Figure \ref{Fig 5}. Extended galaxies form stars slightly faster than compact ones, see more details about the star formation history in Section \ref{sec 4.2}. Galaxies in different mass ranges follow a similar trend. Surprisingly, the sizes of compact galaxies have a small change over the past 10 Gyr, while extended galaxies have grown significantly larger from median size roughly from 1.27 kpc to 4.51 kpc for M1, from 1.43 kpc to 6.45 kpc for M2, from 1.77 kpc to 8.06 kpc for M3, and from 2.20 kpc to 10.31 kpc for M4. At $z = 0$, the overall difference between compact and extended galaxies is 2.5-3 times. Thus, the different evolutionary pathways of extended and compact galaxies are consistent with those of disk- and bulge-dominant galaxies used in \cite[][]{Du2021}.
        
    \begin{figure}[htb!]
            \centering
            \includegraphics[width=0.45\textwidth]{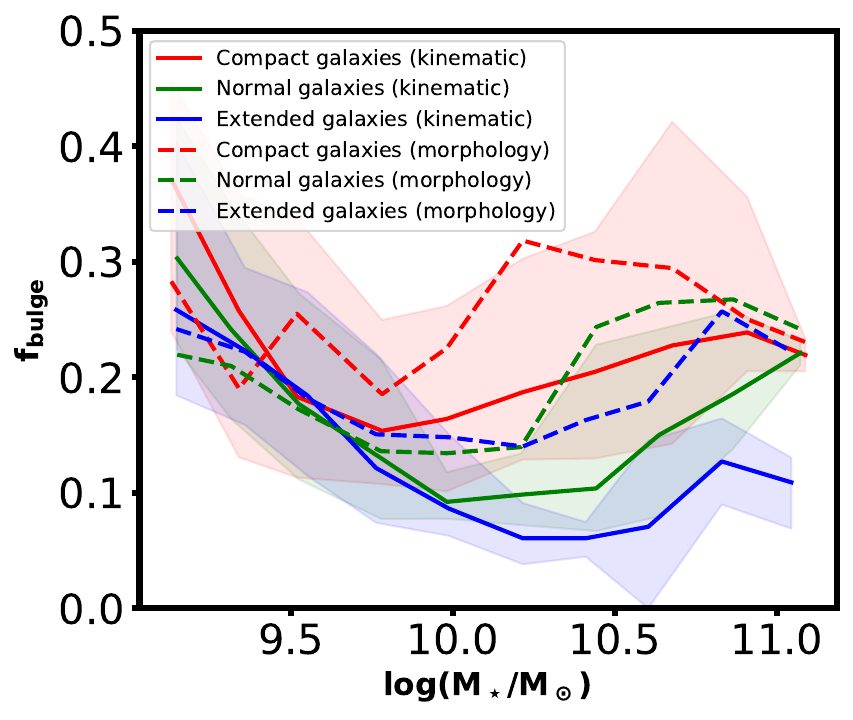}
            \caption{The bulge-to-total mass ratio $f_{\rm bulge}$ as a function of the stellar mass of the galaxy sample we selected. The solid and dashed profiles are results measured by kinematic and morphological methods, respectively. The shaded regions represent the $1\sigma$ envelope of kinematic $f_{\rm bulge}$.}
            \label{Fig 8}
    \end{figure}

        The galaxies with various size and their consequent evolutionary pathways are largely determined by the difference of $j_{\rm h}$. The top panels of Figures \ref{Fig 5} and \ref{Fig 6} show clearly that the host dark matter halos of compact galaxies have similar $M_{\rm h}$, but about a few times lower $j_{\rm h}$ than their extended counterparts. Despite what complex baryonic processes may involve, the difference in specific angular momentum is well maintained as suggested by comparing $j_{\rm gas}$, $j_{\rm cold\, gas}$, and $j_\star$. The cold gas and stars of compact galaxies inherit low angular momenta from their parent dark matter halos, which finally determines the size difference of galaxies, leading to compact and extended evolutionary pathways. This result verifies again that, without effects from external processes, the $j_{\rm h}$ determines the size $R_{1/2}$ of disk galaxies, which is consistent with theoretical arguments \citep[e.g.][]{Fall1980, Mo1998} as well as some hydrodynamic cosmological simulations \citep{Grand2017, Rodriguez-Gomez2022, Yang2023, Du2024}. \citet{Du2024} further showed that disk galaxies in TNG50 follow a tight $j_\star-M_\star$-size scaling relation when mergers are unimportant. Furthermore, the evolutionary path of $j_{\rm gas}$ and $j_{\rm h}$ exhibit remarkable similarities, this indicates that in the absence of baryonic processes, the growth history of angular momentum will not undergo significant changes. It is worth mentioning that the difference of $j_{\rm cold\,gas}$ appears about 3 Gyr later than that of $j_{\rm gas}$. It is reasonable that the gas with high angular momentum cannot spin up the cold gas reservoir immediately due to the pre-existing cold gas with low $j_{\rm cold\,gas}$. And $j_{\star}$ evolves similarly to $j_{\rm cold\,gas}$.

        \begin{figure*}[htb]
            \centering
            \includegraphics[width=0.9\textwidth]{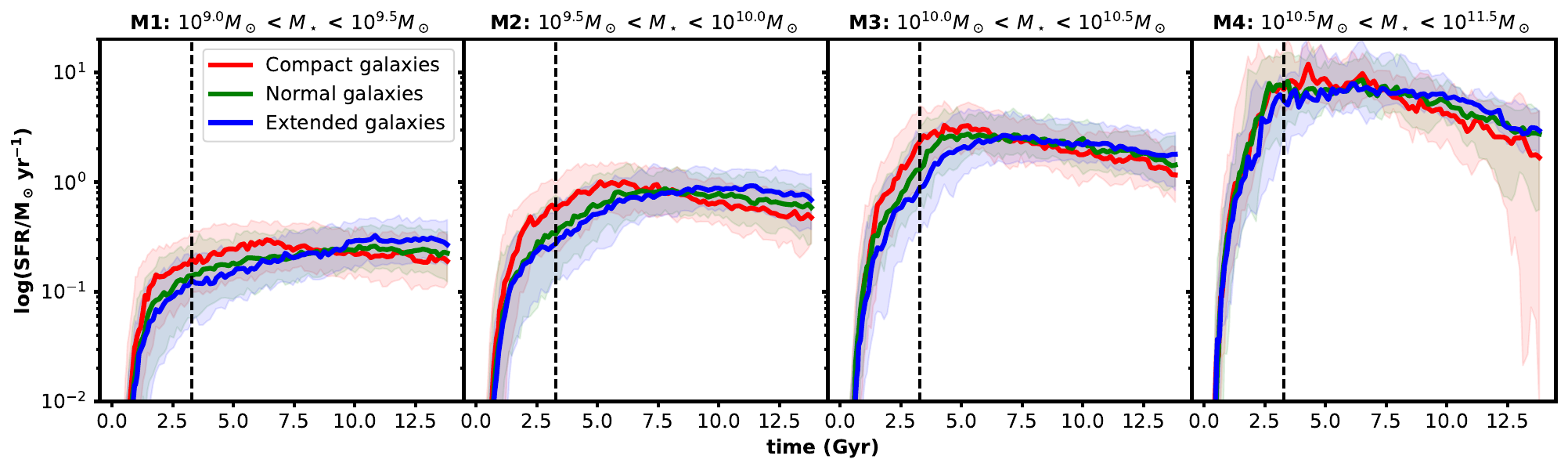}
            \caption{The star formation rate history of compact (red), normal (green), and extended (blue) galaxies we select. The vertical dashed lines mark the time of redshift $z = 2$. The solid profiles and shaded regions represent the median and $1\sigma$ envelopes. From left to right, galaxies are divided into four mass ranges M1-M4.}
            \label{Fig 9}
        \end{figure*}
        \begin{figure}[htb]
            \centering
            \includegraphics[width=0.45\textwidth]{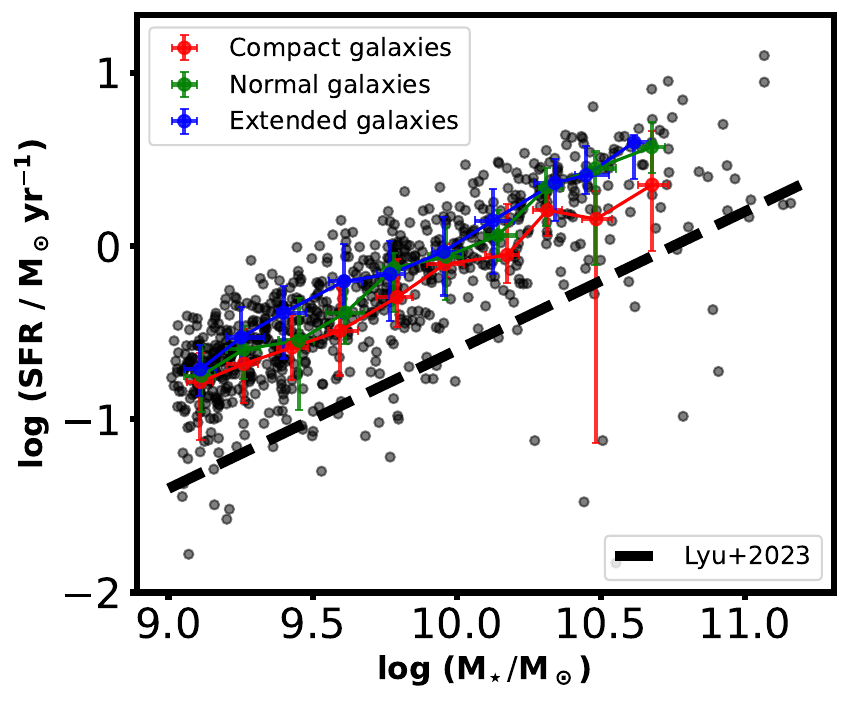}
            \caption{The star formation rate as a function of stellar mass. The colored solid lines represent compact (red), normal (green), and extended (blue) galaxies, respectively. The criterion indicated by the black dashed line, as proposed by \citet{Lyu2023}, is used to differentiate between star-forming and quenched galaxies.}
            \label{Fig 10}
        \end{figure}
        
       Additionally, the difference between compact and extended galaxies is not only due to the difference of $j_{\rm h}$ but also the difference of baryonic fraction. The second row of Figure 5 shows clearly that more compact galaxies accrete significantly less gas than extended ones in the late phase, while their dark matter and stellar components are similar. A similar result is obtained by \cite{Perez_Montano2022}. The $M_{\rm gas}$ difference in three types of galaxies becomes smaller from M1 to M4. It indicates that galaxies of varying sizes are found in different environments, as also shown in Figure \ref{Fig 7}. $M_{\rm b, env}$ and $M_{\rm DM, env}$ are the mass of baryon and dark matter components of the group or cluster that each selected galaxy belongs to. The cumulative distribution of $M_{\rm b, env}/M_{\rm DM, env}$ thus represents the proportion of baryonic mass within the environment where the central galaxies originate. At $z = 0$, the median of $M_{\rm b, env}$/$M_{\rm DM, env}$ reaches $0.140_{-0.022}^{+0.025}$ around extended galaxies, while compact galaxies are generally located at relatively low-dense environment with median $M_{\rm \rm b, env}/M_{\rm DM, env}=0.103$.  It is important to highlight that this deviation is probably attributed to the expansion of dark matter halos in the late phase, as there is no evident distinction at $z=2$. The $p$-value obtained from a Kolmogorov-Smirnov (KS) test comparing compact and extended galaxies at $z=0$ and $z=2$ are nearly 0 and 0.167, respectively. This result suggests that extended galaxies may born in environments where galaxies can accrete more gas. 

        In this section, we show how the angular momenta of dark matter are transferred to galaxies in TNG simulations, which finally determines the galaxy size, as suggested by \citet{Du2022} and \citet{Du2024}. The properties and environment of parent dark matter halos, i.e., nature, play a vital role in the evolutionary history of both compact and extended galaxies. Therefore, the mass-size relation and its scatter are achieved by adjusting the nature of galaxies, including internal physical processes as well as initial $j_{\rm h}$. Moreover, it is well known that the size and mass of galaxies affect significantly also metallicity, star formation rate (SFR), and black hole mass, which are investigated in the next section.

\section{The correlation between the size and other properties of galaxies at $z=0$}\label{sec:4}

    \subsection{More compact galaxies tend to host more massive bulges, but less influence from external processes}\label{sec 4.1}

        The compactness and bulge growth in galaxies are primarily influenced by $j_{\rm h}$. It is thus not surprising that more compact galaxies tend to harbor more massive bulges \citep{Du2021}. Figure \ref{Fig 8} shows $f_{\rm bulge}$ measured by kinematic decomposition adopted from \citet{Du2019, Du2020} (solid profiles) and the result of 1D S\'{e}rsic bulge + exponential disk decomposition in morphology (dashed profiles). It is clear that more compact galaxies with $M_\star>10^{10} M_\odot$ generally tend to host more massive bulges measured by both kinematic and morphological methods. Kinematically-derived $f_{\rm bulge}$ is typically smaller by $\sim 0.1$ than that measured by morphological decomposition. This difference diminishes at lower mass ranges ($M_\star<10^{10} M_\odot$), though the reason is unclear.

        The result above further suggests that galaxies with a lower initial angular momentum are likely to form more massive bulges. Surprisingly, $f_{\rm acc}$ in compact galaxies is smaller than that in more extended cases, as illustrated in the right panel Figure \ref{Fig 3}. In other words, compact galaxies have lower $f_{\rm acc}$ but larger $f_{\rm bulge}$ than extended ones. This suggests that the limited external influences in compact galaxies with small stellar halos may not effectively facilitate bulge growth, contradicting the common belief that mergers are the primary drivers of bulge formation. 
        
    \subsection{Star formation rate is reduced but not quenched in compact galaxies}\label{sec 4.2}
        
        In Figure \ref{Fig 9}, SFR is measured by summing the SFR values of all gas cells within each galaxy at each snapshot. Noteworthy fluctuations in SFR are observed across galaxies of varying sizes. In the early phase, the $j_{\rm cold\,gas}$ values of the three galaxy types are comparable, thereby predominantly shaping the SFR through $M_{\rm cold\,gas}$. However, in the late phase, more extended galaxies generally have higher both $j_{\rm cold\,gas}$ and $M_{\rm cold\,gas}$. Consequently, compact galaxies exhibit greater stellar masses in the early phase. As the evolution progresses, the SFR of extended galaxies slightly exceeds that of compact ones, resulting in similar stellar masses at $z = 0$.
        
        There is no clear signature of ``angular momentum quenching'' in nature. In Figure \ref{Fig 10}, most galaxies (black dots) in this analysis lie above the dashed line that delineates star-forming and quiescent galaxies \citep[e.g.,][]{Lyu2023, Genel2018}. The median SFR values at $z=0$ for compact, normal, and extended galaxies are denoted by red, green, and blue dots with accompanying error bars, respectively. While compact galaxies demonstrate a slightly lower SFR compared to their extended counterparts, the scatter in data points is considerable. The presence of a limited number of quenched galaxies is likely instantaneous or a consequence of AGN feedback stemming from their comparatively more massive black holes.  These results suggest that the so-called ``angular momentum quenching,'' as proposed by \citet{Peng2020}, lacks a clear manifestation when external influences are not taken into account.
        \cite{Wang2022} and \cite{Lu2022} demonstrated that external mechanisms inject angular momentum into the circumgalactic medium, influencing or potentially halting star formation. Consequently, the increase in angular momentum may be one consequence of mergers rather than a direct approach to quench galaxies. In conclusion, our results show that variations in angular momentum obtained by gas accretion from the cosmic web are not enough to quench galaxies, highlighting the necessity of external physical processes.

        \begin{figure*}[htb]
            \centering
            \includegraphics[width=1\textwidth]{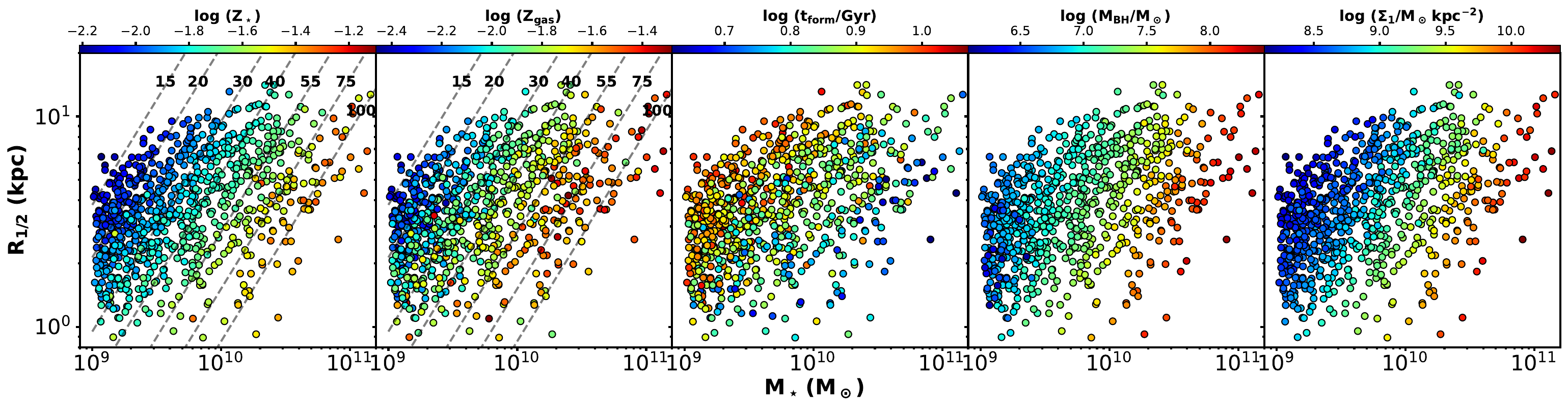}
            \caption{The stellar mass-size relation for the galaxies in the TNG50 simulation at $z = 0$. From left to right, the data is color-coded by the stellar metallicity ($Z_{\star}$), gas metallicity ($Z_{\rm gas}$), galaxy formation time ($t_{\rm form}$), black hole mass ($M_{\rm BH}$), and stellar density measure within central 1 kpc in the cylindrical coordinate ($\Sigma_1\ (2D)$). Diagonal dashed lines represent lines of constant velocity dispersion, calculated using the formula $\sigma_e = \sqrt{\frac{GM_{\star}}{5R_{1/2}}}$.}
            \label{Fig 11}
        \end{figure*}

    \subsection{Mass-size-$X$ scaling relations: implications for formation time, metallicity, central concentration, and black hole mass}\label{sec 4.3}

        Both the size and stellar mass of galaxies correlate tightly with the other properties of galaxies. We examine the scaling relations that are determined by the mass and size in Figure \ref{Fig 11}. A power-law relationship can be used to describe such scaling relations, as follows
        \begin{equation}\label{equation 3}
            {\rm log} X = A\ {\rm log} \frac{M_\star/M_\odot}{(R_{\rm 1/2}/{\rm kpc})^\alpha} + B, 
        \end{equation}
        where the index $\alpha$ measures the extent to which properties $X$ are correlated with both $M_\star$ and $R_{\rm 1/2}$. It is evident that the metallicity in both gas and stars shows a synchronized decrease on the $M_\star$--$R_{\rm 1/2}$ diagram. Therefore, we conduct a surface fitting in the three-dimensional (3D) space. The results of the fitting for galaxy formation time ($t_{\rm form}$), metallicity, and black hole mass are presented in Table \ref{tabel 1}, offering an empirical $M_\star$--$R_{\rm 1/2}$--$X$ relationship derived from the nature of galaxies in TNG simulations.

        \subsubsection{Metallicity}

        We investigate how the stellar and gas metallicity of galaxies are related to their mass and size in our sample, as shown in the first and second panels of Figure \ref{Fig 11}. Our analysis reveals that galaxies with smaller sizes $R_{1/2}$ tend to have higher metallicity when they have the same mass. Similarly, galaxies with the same size $R_{1/2}$ exhibit higher metallicity as their mass increases, in line with the results of \cite{Vaughan2022}. Hence, the metallicity of a galaxy is not only linked to its mass but also to its size. This observation aligns with previous findings and theoretical predictions \citep{Barone2018, Barone2020, D'Eugenio2018}, which indicate a stronger correlation between the metallicity of stars $Z_{\star} = M_{\rm star, Z}/M_{\rm star, total}$ and gas $Z_{\rm gas} = M_{\rm gas, Z}/M_{\rm gas, total}$ and the mass and size of galaxies, rather than just their masses.
        
        The underlying physical mechanisms driving this relationship are not yet fully understood. It is commonly thought to be connected to the gravitational potential well of the galaxy, as galaxies with deeper potential wells can retain more metal-rich gas due to their higher escape velocities \citep{Franx1990, Emsellem1996, Scott2009, Scott2017}. The fitting parameter $\alpha$ for the metallicity of stars and gas, as shown in Table \ref{tabel 1}, is approximately 1. The relationship $\frac{GM_\star}{R_{\rm 1/2}} \propto \Phi \propto \sigma_{e}^2$ suggests that gravitational potential or velocity dispersion can be used to predict the metallicity of galaxies, which is consistent with previous studies' findings \citep{Barone2018, Barone2020, Barone2022, Vaughan2022}. This result further suggests that the dark matter halo may influence the metal enrichment of disk galaxies by determining their sizes and stellar masses.
        
        \subsubsection{Galaxy formation time, the growth of central concentration, and super-massive black holes}
        
        Compact galaxies form earlier than their extended counterparts, as indicated in Section \ref{sec 3.1}. The galaxy formation time $t_{\rm form}$ is the time when the stellar mass $M_\star$ reaches half its value at $z = 0$. The mass-size relation color-coded by $t_{\rm form}$ is shown in the third panel of figure \ref{Fig 11}, clearly illustrating that compact galaxies form earlier. This relationship can also be well described by Equation \ref{equation 3}. The fitting results give ${\rm log}t_{\rm form}\simeq -0.113\ {\rm log} [M_\star/(R_{\rm 1/2})^{1.638}]+1.873$ (Table \ref{tabel 1}).
        
        Moreover, more compact galaxies tend to form earlier and grow more massive bulges, as demonstrated in Section \ref{sec 4.1}. The inflow of cold gas into the galaxy's center potentially also promotes the fast growth of the central concentration and the black hole's mass \citep{Grand2017, Genel2018, Rodriguez-Gomez2022}. It's not surprising that compact galaxies tend to harbor more massive black holes, partly because more gas can reach the galaxies' central regions. Another factor is the impact of AGN feedback on galactic angular momentum, which primarily occurs in later stages, suppressing late-time gas accretion in galaxies with more massive black holes. Since gas accreted late typically has higher specific angular momentum, its absence could explain why galaxies with larger black hole masses exhibit lower $j_\star$ values \citep{Grand2017, Rodriguez-Gomez2022}.

        The growth of the bulge can also be quantified by central stellar densities measured within 1 kpc from the galactic center in either cylindrical (2D) or polar (3D) coordinate $\Sigma_1$. In Figure \ref{Fig 11}, we can see the mass of central supermassive black holes follow a similar parallel manner to $\Sigma_1$. More massive galaxies with smaller sizes tend to have larger central densities as well as larger black hole masses. Empirical correlations are given in Table \ref{tabel 1} via fitting the parameters of Equation \ref{equation 3}. The deviation of the $M_{\rm BH}$ scaling relation from observational findings ${\rm log} M_{\rm BH} \propto {\rm log} [M_\star/(R_{\rm 1/2})^{1.073}]$ in \citet{Krajnovi'c2018} is not unexpected, given the significant measurement uncertainties and the possibility of inaccurate subgrid-physics in TNG simulations. 
        
        It is worth mentioning that the increase in bulge mass and central concentration $\Sigma_1$ has been proposed as a mechanism for suppressing star formation in galaxies \citep[e.g.][]{Martig2009, Fang2013, Lang2014}. While compact galaxies do exhibit a somewhat lower SFR, it remains uncertain whether the growth of bulges directly leads to SFR suppression in disk galaxies. Alternatively, the decrease in SFR is likely to be linked to the depletion of gas during the evolution of compact galaxies in the late phase.
        
        We do not anticipate the TNG simulations to accurately reproduce all properties and scaling relations of galaxies that we discuss here. Especially, the resolution of TNG simulations cannot sufficiently resolve the central part of galaxies. However, the TNG simulation can provide a physical vantage point for comprehending the more intricate physical mechanisms derived from the fundamental properties (mass and size) of galaxies in nature. In future simulations, researchers can examine the alterations in these correlations by modifying the physical models.
        
        \begin{table}[]
            \centering
            \scriptsize
            \caption{Best-fit parameters for the mass-size-$X$ scaling relations of central disk galaxies in TNG}
            \begin{tabular}{lcccc}
                \hline
                Galaxy properties ($X$) & Slope ($A$) & $\alpha$ & Intercept ($B$) & Dispersion \\
                \hline
                $Z_\star$ & 0.374$\pm$0.004 & 1.099$\pm$0.021 & -5.239$\pm$0.040 & 0.054 \\
                $Z_{\rm gas}$ & 0.490$\pm$0.011 & 0.937$\pm$0.038 & -6.417$\pm$0.098 & 0.129 \\
                $t_{\rm form}$ & -0.113$\pm$0.005 & 1.638$\pm$0.087 & 1.873$\pm$0.047 & 0.064 \\
                $M_{\rm BH}$ & 0.936$\pm$0.010 & 0.613$\pm$0.019 & -1.619$\pm$0.089 & 0.120 \\
                $\Sigma_{1} (2D)$ & 1.017$\pm$0.008 & 0.829$\pm$0.015 & -0.462$\pm$0.078 & 0.106 \\
                $\Sigma_{1} (3D)$ & 1.051$\pm$0.010 & 0.871$\pm$0.017 & -0.821$\pm$0.095 & 0.129 \\
                \hline
        \end{tabular}
            \label{tabel 1}
            \raggedright
            \textbf{Note.} The slope (A), parameter $\alpha$, and intercept (B) values for star metallicity $Z_\star$, gas metallicity $Z_{\rm gas}$, galaxy formation time $t_{\rm form}$, black hole mass $M_{\rm BH}$, and stellar density within 1 kpc $\Sigma_{1} (2D/3D)$ were obtained through the optimal fitting parameters in Equation \ref{equation 3}. The last column is the dispersion (i.e., the standard deviation of the residuals) of the best-fitted Equation \ref{equation 3}.
        \end{table}

\section{Summary}\label{sec:5}

    In this study, we use the data from the TNG50--1 simulation to categorize central disk galaxies into three typical morphologies: compact, normal, and extended. We thoroughly investigated the growth histories of each galaxy type, focusing on their various components (e.g., dark matter, star, gas). Additionally, we explored the connection between galaxy properties and their masses and sizes driven by nature, providing insights through an analysis of their growth history. Our primary results can be summarized as follows:
    \begin{itemize}
        \item In the early phase $z>2$, more compact galaxies form earlier, while all galaxies exhibit similar specific angular momentum $j_\star$ and size $R_{1/2}$, typically around 1 kpc, aligning with observational data. Therefore, the slope of the mass-size relation is flat. Additionally, we find that compact galaxies possess a greater amount of cold gas, leading to earlier stellar mass accumulation.
        
        \item In the later phase, there are significant discrepancies in $j_{\rm h}$ between compact and extended galaxies, with no substantial variations in $M_{\rm h}$. Despite the involvement of complex baryonic physical mechanisms, galaxies retain the specific angular momentum of their parent dark matter halos, which dictates the rate of growth in size $R_{1/2}$. Galaxies with higher $j_\star$ experience a quicker increase in size, resulting in a steeper slope in the mass-size relationship. The presence of different $j_\star$ values for the same mass contributes to the dispersion in the mass-size relation.

        \item All types of galaxies are star-forming. There is no clear signature of ``angular momentum quenching'' in galaxies we selected where no external processes have been considered.

        \item More compact galaxies retain more metals which is consistent with observations and the theoretical prediction that the gravitational potential well determines the metallicity enrichment. It gives an empirical scaling relation that ${\rm log} Z_{\rm gas}$ and ${\rm log}Z_{\star}$ are proportional to $\ {\rm log} (M_\star/R_{\rm 1/2})$. Additionally, for galaxies of the same size, those with a larger mass formed earlier and have higher metallicity. 
        
        \item A more compact galaxy generally has faster cold gas accretion into its central regions, which accelerates the mass growth of both the bulge and black holes. We propose an empirical scaling relation, i.e., ${\rm log} M_{\rm BH}\simeq 0.936\ {\rm log} [M_\star/(R_{\rm 1/2})^{0.613}]-1.619$. Additionally, more compact galaxies also tend to develop a larger bulge and central concentration at a faster rate.
        
    \end{itemize}

These findings confirm that the characteristics of galaxies are strongly linked to their associated dark matter halos. In situations where external factors are not important, the stellar mass and size of galaxies play a significant role in shaping their various properties.
    
\begin{acknowledgements}
The authors acknowledge the support by the Natural Science Foundation of Xiamen, China (No. 3502Z202372006), the
Fundamental Research Funds for the Central Universities (No. 20720230015), and the Science Fund for Creative Research Groups of the National Science Foundation (NSFC) of China (No. 12221003). 
L.C.H. was supported by the National Science Foundation of China (11991052, 12233001), the National Key R\&D Program of China (2022YFF0503401), and the China Manned Space Project (CMS-CSST-2021-A04, CMS-CSST-2021-A06).
S.L. acknowledges the support by the NSFC grant (No. 11988101) and the K. C. Wong Education Foundation.
The TNG50 simulation used in this work, one of the flagship runs of the IllustrisTNG project, has been run on the HazelHen Cray XC40-system at the High-Performance Computing Center Stuttgart as part of project GCS-ILLU of the Gauss centers for Supercomputing (GCS). This work is also strongly supported by the Computing Center in Xi'an. 
\end{acknowledgements}

\bibliographystyle{aa} 
\bibliography{ref} 
\end{document}